\documentclass[a4paper]{spie}  %>>> use this instead for A4 paper
\usepackage[]{graphicx}
\usepackage{amsmath}
\usepackage{amssymb}
\usepackage{xspace}
\usepackage{times}

\newcommand{\degree}{\ensuremath{^\circ}\xspace}
\newcommand{\gagg}{\ensuremath{g_{a\gamma}}\xspace}
\newcommand{\fefifty}{${}^{55}{\rm Fe}$\xspace}
\newcommand{\hefour}{\ensuremath{{}^{4}{\rm He}}\xspace}
\newcommand{\hethree}{${}^{3}{\rm He}$\xspace}
\newcommand{\pbtwoten}{\ensuremath{{}^{210}{\rm Pb}}\xspace}
\newcommand{\Lsun}{\ensuremath{L_\odot}\xspace}

\title{pn-CCDs in a Low-Background Environment: Detector Background of the CAST X-ray Telescope}

\author{M. Kuster \supit{a,d}, S. Cebri\'an\supit{b}, A.
  Rodr\'iquez\supit{b}, R. Kotthaus \supit{c}, H. Br\"auninger \supit{d},
  J. Franz \supit{e}, P.  Friedrich \supit{d}, R. Hartmann \supit{f}, D.
  Kang \supit{e}, G. Lutz \supit{c}, L. Str\"uder\supit{d}
  \skiplinehalf 
  \supit{a}Technische Universit\"at Darmstadt, Schlossgartenstr. 9,
  64289 Darmstadt, Germany\\
  \supit{b}Instituto de F\'isica Nuclear y Altas Energ\'ias, Universidad de
  Zaragoza, Zaragoza, Spain\\
  \supit{c}Max-Planck-Institut f\"ur Physik, F\"ohringer Ring 6, 80805
  M\"unchen, Germany\\
  \supit{d}Max-Planck-Institut f\"ur extraterrestrische Physik,
  Giessenbachstr., 85748 Garching, Germany\\ 
  \supit{e}Universit\"at Freiburg -- Physikalisches Institut,
  Herrman-Herder-Str. 3, 79104 Freiburg, Germany\\
  \supit{f}PNSensor GmbH, R\"omerstr. 28, 80803 M\"unchen, Germany}

\authorinfo{Send correspondence to M. Kuster: E-mail: kuster@hll.mpg.de,
  Phone: +49 (0)6151 16-2321}

\begin{document} 
\maketitle 

%%%%%%%%%%%%%%%%%%%%%%%%%%%%%%%%%%%%%%%%%%%%%%%%%%%%%%%%%%%%% 
\begin{abstract}
  The CAST experiment at CERN (European Organization of Nuclear Research)
  searches for axions from the sun. The axion is a pseudoscalar particle
  that was motivated by theory thirty years ago, with the intention to
  solve the strong CP problem. Together with the neutralino, the axion is
  one of the most promising dark matter candidates. The CAST experiment has
  been taking data during the last two years, setting an upper limit on the
  coupling of axions to photons more restrictive than from any other solar
  axion search in the mass range below $10^{-1}\,\text{eV}$. In 2005 CAST
  will enter a new experimental phase extending the sensitivity of the
  experiment to higher axion masses.
  
  The CAST experiment strongly profits from technology developed for high
  energy physics and for X-ray astronomy: A superconducting prototype LHC
  magnet is used to convert potential axions to detectable X-rays in the
  $1$--$10\,\text{keV}$ range via the inverse Primakoff effect. The most
  sensitive detector system of CAST is a spin-off from space technology, a
  Wolter I type X-ray optics in combination with a prototype pn-CCD
  developed for ESA's XMM-Newton mission. As in other rare event searches,
  background suppression and a thorough shielding concept is essential to
  improve the sensitivity of the experiment to the best possible. In this
  context CAST offers the opportunity to study the background of pn-CCDs
  and its long term behavior in a terrestrial environment with possible
  implications for future space applications. We will present a systematic
  study of the detector background of the pn-CCD of CAST based on the data
  acquired since 2002 including preliminary results of our background
  simulations.
\end{abstract}

%>>>> Include a list of keywords after the abstract 

%%%%%%%%%%%%%%%%%%%%%%%%%%%%%%%%%%%%%%%%%%%%%%%%%%%%%%%%%%%%%
\section{Introduction}
\label{sect:intro}
The most sensitive axion helioscope in operation, the CERN axion solar
telescope -- CAST, aiming to discover the elusive particle axion, has
recently finished its first data taking period. Although no signal over
background could be detected by any of the detectors of CAST during the
first period of operation, the results from data taken during 2003 allow to
improve the existing upper limit on the coupling of axions to photons to
$\gagg<1.16\times10^{-10}\,\text{GeV}^{-1}$ (for $m_{\text{a}}<
0.02\,\text{eV}$, see Fig.~\ref{fig:axion-exclusion})\cite{zioutas:04d}.
The axion as a particle is a direct
consequence\cite{weinberg:78a,wilczek:78a} of the Peccei-Quinn
mechanism\cite{peccei:77a} proposed in 1977 to solve the still existing
strong CP problem. The strong CP problem describes the fact that CP
violation in strong interactions seems not to be realized in nature, albeit
the QCD Lagrangian density contains CP violating terms. This inconsistency
between theory and experiment becomes apparent by the fact, that the best
experimental limit for the electric dipole moment of the neutron
corresponds to $\approx 10^{-9}$ of the range allowed within QCD.  As
pointed out by several authors, e.g. G.  Raffelt\cite{raffelt:05a}, axions
of a mass in the sub eV range would also be a viable hot and cold dark
matter candidate.

If the axion exists, it would couple to two photons with the strength given
by the coupling constant $\gagg$. This coupling would allow the production
of axions inside the hot plasma of stars via the so called Primakoff effect
($\gamma\gamma'\to\text{a}$), where photons interact with the coulomb field
of plasma particles and are converted into axions. The expected axion flux
emitted by the sun is $\Phi_{\text{a}}=g_{10}^2\cdot
3.77\times10^{11}\,\text{axions}\,\text{cm}^{-2}\,\text{sec}^{-1}$
corresponding to an axion luminosity of $L_{\text{a}}=g_{10}^2\cdot
1.7\times10^{-3}\Lsun$ with
$g_{10}=\gagg\times 10^{10}\,\text{GeV}$\cite{raffelt:05a}. The
energy spectrum of the axions would reflect the thermal energy spectrum of
the photons in the core of the sun with a mean energy of $\approx
4.2\,\text{keV}$. To detect axions on Earth, Sikivie \cite{sikivie:83a}
proposed an experimental approach called the ``axion helioscope'' which the
CAST experiment is based on:
\begin{figure}
  \centerline{\includegraphics[width=0.65\textwidth]{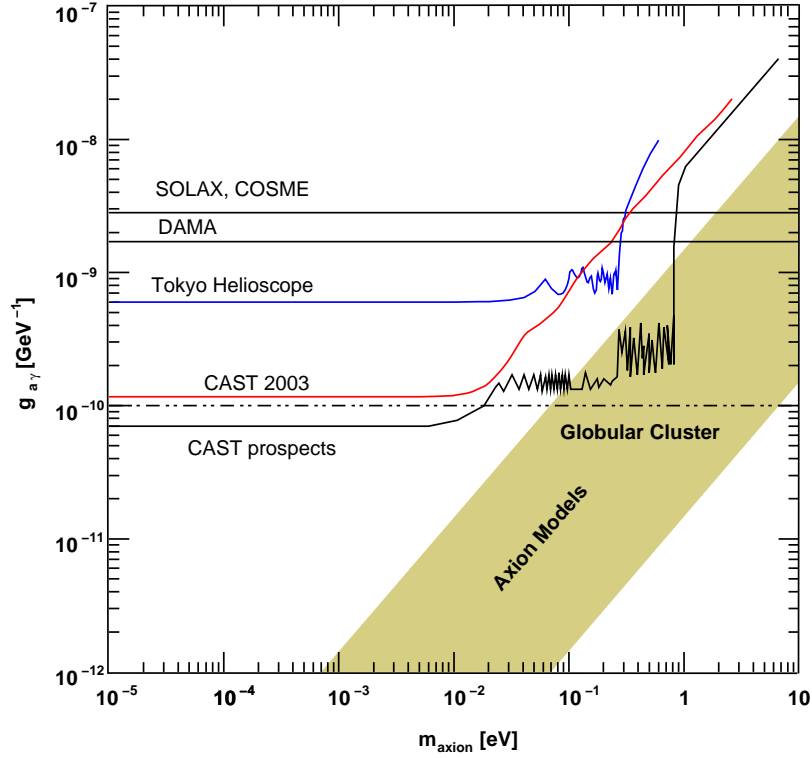}}
  \caption 
  {\label{fig:axion-exclusion} Axion mass $m_{\text{a}}$ versus the axion
    to photon coupling constant \gagg parameter space. The red line marks
    the upper limit on \gagg derived from CAST data taken in 2003. The
    shaded area labeled ``Axion Models'' marks the region favored by
    theoretical axion models. Results from the previous experiments SOLAX,
    COSME, DAMA, and the Tokyo Helioscope are shown for comparison. The
    dash dotted line labeled ``Globular Clusters'' marks the limit on \gagg
    derived from astrophysical considerations\cite{zioutas:04d}.}
\end{figure} 
in the presence of a transverse magnetic field solar axions could be
converted back to observable X-rays via the time reversed Primakoff effect.
In CAST we use a $9\,\text{Tesla}$ superconducting dipole magnet for this
purpose, providing a homogeneous transversal magnetic field inside two
$9.26\,\text{m}$ long magnet tubes with a diameter of $43\,\text{mm}$ each.
The magnet is supported by a movable platform that allows to point the
magnet to the sun for $3\,\text{h}$ per day ($1.5\,\text{h}$ during sun
rise and sun set). At each end of the magnet three X-ray detectors (a TPC
covering two magnet bores, a Micromegas detector, and an X-ray telescope)
are looking for an excess signal over background in order to detect a
possible axion signal.  The probability for an axion to be converted to an
X-ray photon in the magnetic field depends on the magnetic field strength
$B$, the length of the magnetic field $L$, the momentum transfer from the
axion to the photon $|\vec{q}|=|\frac{m_{\text{a}}^2-m_{\gamma}^2}{2E}|$,
the absorption coefficient $\Gamma$ of the medium inside the conversion
volume, and the axion to photon coupling constant \gagg according
to\cite{bibber:89a}:
\begin{equation}
  P_{a\rightarrow\gamma}=\left(\frac{B\gagg}{2}\right)^2
        \frac{1}{q^2+\Gamma^2/4} \left[ 1+e^{-\Gamma L}-2e^{-\Gamma L/2}\cos(qL)\right]
\end{equation}
During the first phase of operation of CAST in 2003--2004 (Phase I) the
conversion volume of the magnet was evacuated (i.e. $\Gamma=0$ and
$m_{\gamma}=0$).  Then the conversion probability depending on the axion
mass remains constant for a given \gagg and given magnet parameters $B$ and
$L$, as long as the momentum transfer from the axion to the outgoing photon
is negligible or in other words $qL \ll \pi$.  The expected integrated
X-ray photon flux in the energy range of $1$--$7\,\text{keV}$ is then
$\Phi_{\gamma}=0.51\,g_{10}^4\left(\frac{L}{9.26\,\text{m}}\right)^2\left(\frac{B}{9\,\text{T}}\right)^2\,\text{photons}\,\text{cm}^{-2}\,\text{d}^{-1}$.
For higher axion masses, $m_\text{a} \gtrsim 0.02\,\text{eV}$, the
conversion probability rapidly drops and limits the sensitivity of the
experiment with the magnet pipes being evacuated. For the end of 2005 it is
planned to fill the magnet bore with a refractive medium (\hefour first and
then \hethree), such that the photon acquires an ``effective'' mass and the
axion to photon momentum mismatch can be overcome (CAST Phase~II).  During
Phase~II, CAST has the potential to probe regions in the
$\gagg$--$m_{\text{a}}$ parameter space that were not reachable for any
other experiment so far. Especially the parameter range that is favored by
theoretical axion models (region labeled ``axion models'' in
Fig.~\ref{fig:axion-exclusion}) can experimentally be probed for the first
time with CAST.

In 2003 the three detector systems of CAST have taken useful data for more
than $260\,\text{h}$ during axion sensitive conditions. In addition, more
than $1233.5\,\text{h}$ of detector background data were acquired in 2003
with the X-ray telescope under different operating conditions. This data
base, plus $179.4\,\text{h}$ of tracking data and $1723.5\,\text{h}$ of
background data from the data taking runs in 2004, allow a systematic study
of the performance and of the background observed with the most sensitive
detector system of CAST, the X-ray telescope in conjunction with the pn-CCD
detector.

\section{The CAST X-ray Telescope and pn-CCD detector}
Since the axion to photon conversion inside the magnet tubes conserves the
momentum of the incoming axion, the resulting X-rays would leave the magnet
bore as a nearly parallel beam. The divergence is given by the angular size
of the magnet aperture fully covering the axion producing region of the
sun. The X-ray flux can then either be observed directly with a detector
mounted to the magnet bore, like it is the case for the Micromegas and TPC
detectors, or it can be focused with an X-ray optics onto a focal plane
detector with a high spatial resolution. The advantage of the latter
approach is twofold: in case of a positive signal a telescope system would
allow to aquire an image of the axion distribution in the core of the sun
and the focusing of X-rays coming from the magnet bore from an area of
$14.5\,\text{cm}^2$ to a small spot with an area of $\approx 6.4
\,\text{mm}^2$ improves the signal to background ratio and thus the
sensitivity of the experiment, significantly.  In CAST we realized such an
imaging system, consisting of a prototype Wolter~I type X-ray telescope
developed for the German X-ray satellite ABRIXAS
\cite{altmann:98a,egle:98a} and a pn-CCD detector similar to the fully
depleted EPIC-pn focal plane detector of
XMM-Newton\cite{kuster:04d,lutz:04a}. A detailed summary on the performance
and characteristics of the pn-CCD detector of XMM-Newton can be found in
Ref.~\cite{strueder:01a} and references therein. The backside illuminated
CCD chip of CAST is operated at a temperature stabilized at -130\degree C
and has a sensitive area of $2.88\,\text{cm}^2$ divided into $200\times64$
square pixels, providing a quantum efficiency close to unity between
$1$--$7\,\text{keV}$.
\begin{figure}
  \centerline{\includegraphics[width=0.9\textwidth]{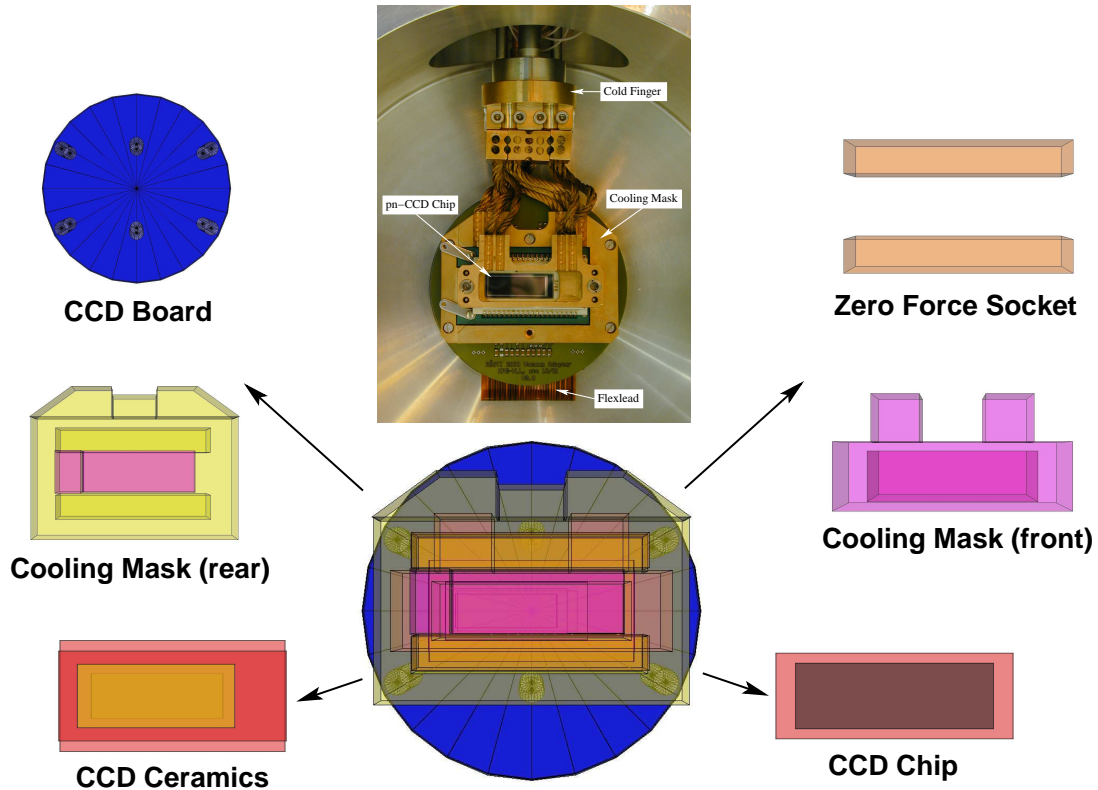}}
  \caption{\label{fig:pnccdchip} Top Center: The pn-CCD chip (black part in
    the center) with the gold plated cooling mask which is connected to a
    cold finger of a Stirling cooler device. The individual detector
    components as modeled in GEANT4 are shown below and to the left and
    right. Due to the limited time for R\&D these components were tested
    for their natural radioactivity in the Canafranc underground
    laboratory\cite{kuster:04d}, but not built from selected and radio-pure
    materials.}
\end{figure}

Fig.~\ref{fig:pnccdchip} is a picture showing the interior of the detector,
the CCD chip, the cooling mask, and the electronic components necessary to
operate the CCD. In addition the different detector components are shown as
schematic drawings as represented in the GEANT4 model we use for background
simulations. An electronics board is fixed to the rear side of the CCD,
carrying electronic components soldered to the board. The CCD chip itself
is glued onto an aluminum oxide substrate which is sandwiched by a Cu
cooling mask from the front and the back, providing a good thermal coupling
to the cold finger. The CCD chip is surrounded by a passive shield inside
the vacuum vessel build of a $10$--$40\,\text{mm}$ thick copper box made of
low activity, oxygen free copper. The copper box is surrounded by a
$22\,\text{mm}$ thick lead layer free of \pbtwoten, followed by additional
$2.5\,\text{cm}$ of lead outside the vacuum vessel. The detector in this
final configuration was installed at the CAST magnet in spring 2004 before
the 2004 data taking runs.

\subsection{Detector Performance}
\begin{figure}
  \vspace{-1.4cm}
  \centerline{\includegraphics[width=0.93\textwidth]{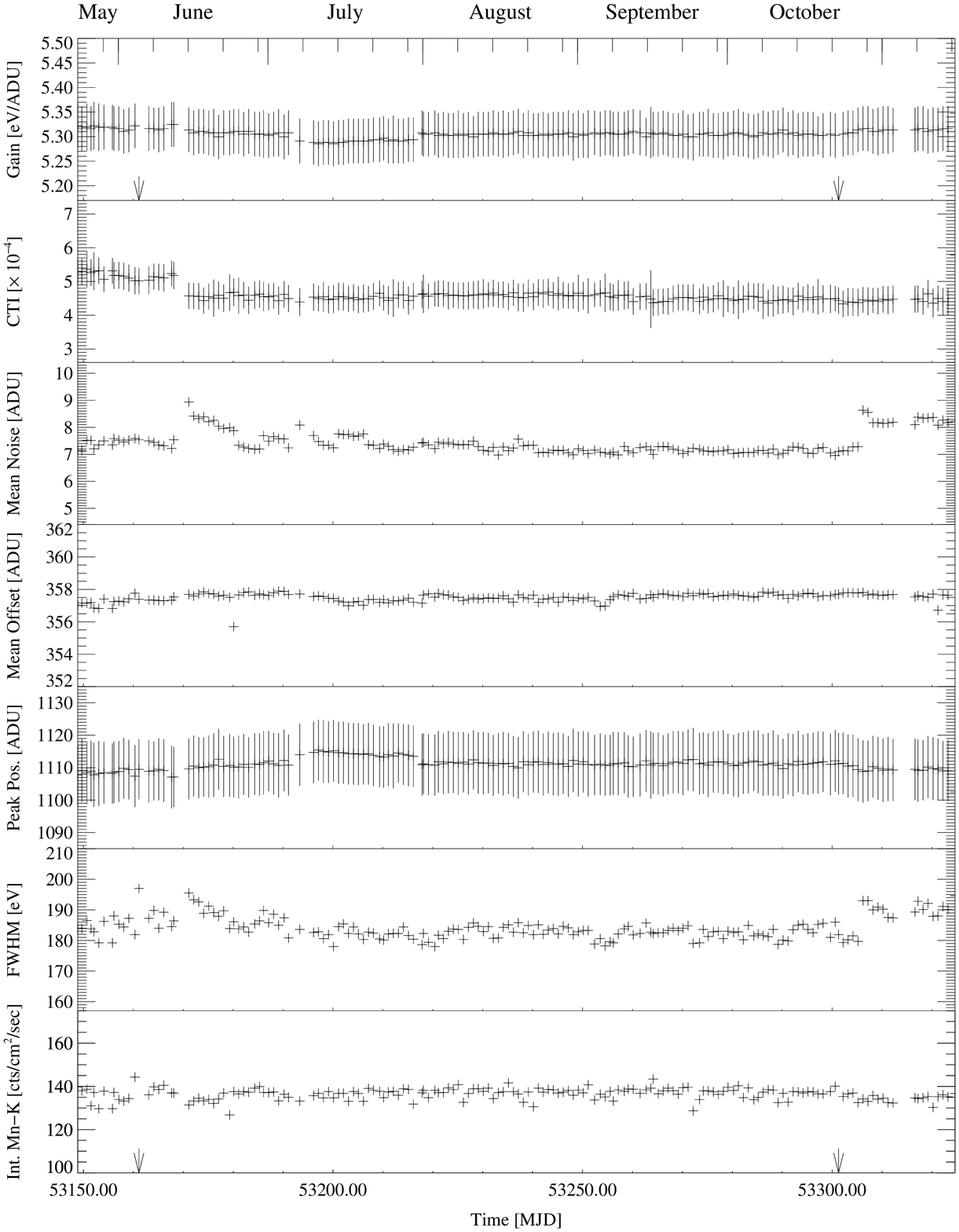}}
  \caption 
  { \label{fig:performace-summary} Performance of the pn-CCD Detector of
    CAST during the data taking period of 2004. From top to bottom: The
    amplification (ADU to keV conversion), charge transfer inefficiency
    (CTI), mean noise averaged over all pixels, the mean offset averaged
    over all pixels, the peak position of the $\text{Mn-K}_\alpha$ line of
    the calibration source, the energy resolution, and the intensity of the
    $\text{Mn-K}_\alpha$ are shown. None of the detector parameters shows a
    significant variation over the time of operation.}
\end{figure} 
The pn-CCD detector was operated almost continuously during the data taking
periods of CAST in 2003 and 2004. In total we accumulated more than
$2960\,\text{h}$ of useful background data during both years, including
daily calibration measurements with an \fefifty source using a flat field
illumination. This data set allows us to study parameters defining the
performance of the detector in a low background environment over a period
of 2 years. The most important detector parameters, the detector noise
averaged over all pixels, the charge transfer inefficiency (CTI), the
amplification (i.e. the ADU to keV conversion), and the dark current
(offset) averaged over all pixels are summarized in
Fig.~\ref{fig:performace-summary} for the data taking period of 2004.
Similar results could be derived for the 2003 data. All parameters except
the energy resolution given in FWHM of the $\text{Mn-K}_{\alpha}$ line and
the detector noise show no significant variation over the entire period of
operation. The increase of detector noise (mid June and at the end of
October) is correlated to an overall increase of noise in the experimental
area of CAST. This change of noise is apparent from the width of the
$\text{Mn-K}_{\alpha}$ given as FWHM as well. A degradation of, e.g. the
CTI due to radiation damage defects like it is the case for the CCD
detectors of XMM-Newton operated in orbit, is not expected in a ground
based environment.

\section{The Detector Background}
\begin{figure}
  \begin{center}
    \begin{tabular}{c}
      \includegraphics[width=0.88\textwidth]{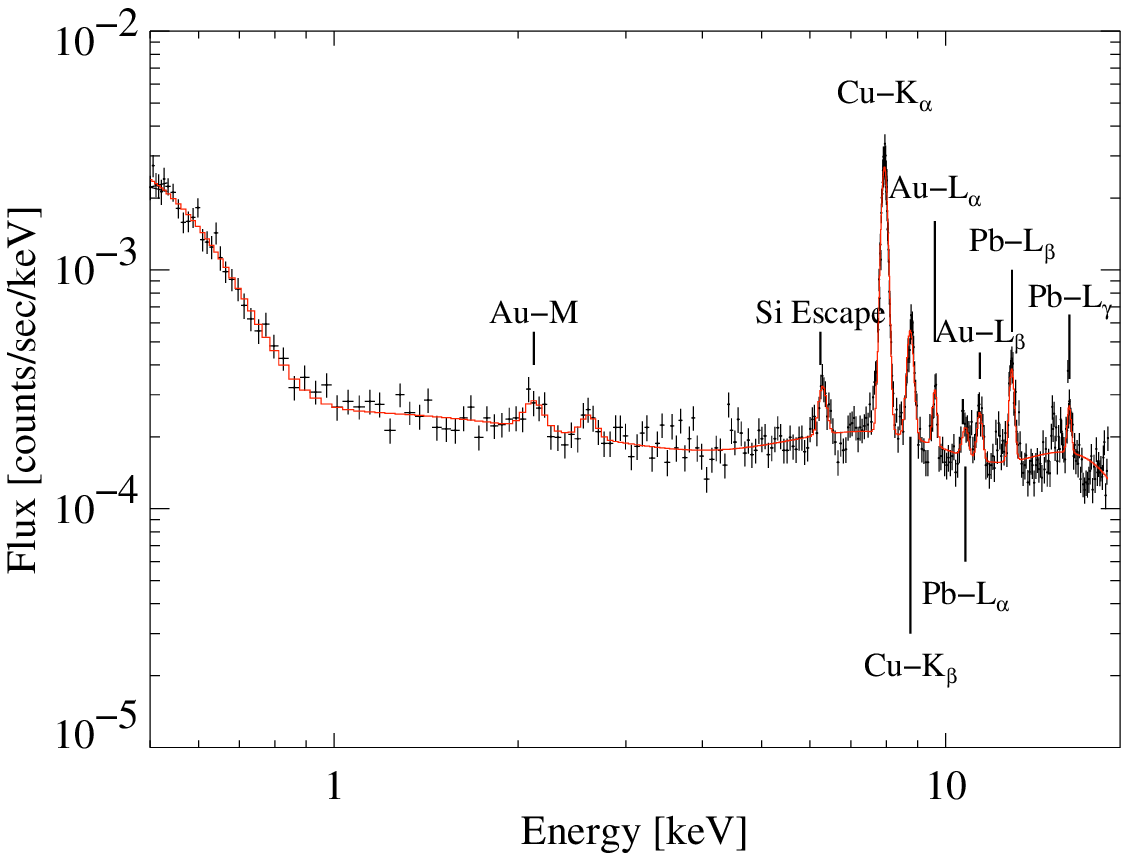}
    \end{tabular}
  \end{center}
  \vspace{-0.3cm}
  \caption 
  { \label{fig:bgrdspec} Background spectrum observed with the pn-CCD
    detector in the CAST environment. The overall background is composed of
    contributions from fluorescent emission of Pb, Cu, and Au on top of a
    continuum spectrum. The peak at energies $E<1\,\text{keV}$ corresponds
    to low energy noise. The X-rays from axion to photon conversion are
    expected to be thermally distributed between $1.0\,\text{keV}$ and
    $8\,\text{keV}$.}
\end{figure} 
The main contributions to the overall background of the CAST pn-CCD
detector is external background induced by cosmic rays, gamma rays, and
radioactive impurities of structural magnet materials. In addition the
intrinsic detector background due radioactive impurities in the detector
materials, or the shape of the response function of the pn-CCD can
significantly contribute to the total observed background.
\begin{figure}
  \begin{minipage}{0.49\textwidth}
    \includegraphics[width=1.0\textwidth]{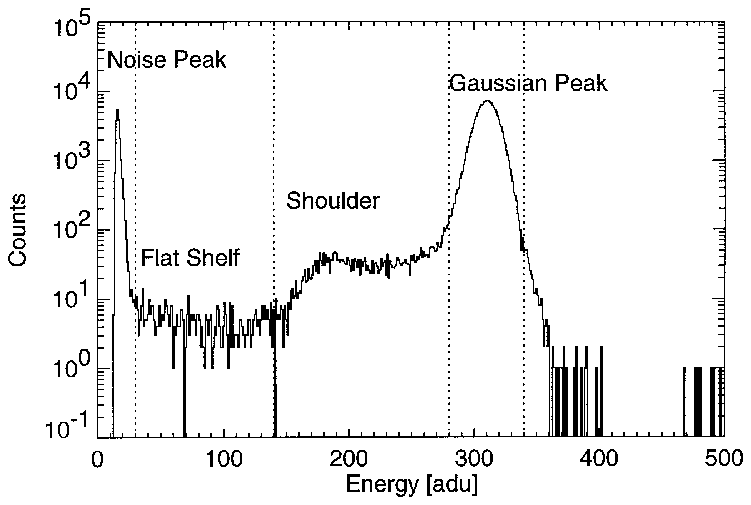}
  \end{minipage}
  \begin{minipage}{0.49\textwidth}
    \includegraphics[width=1.0\textwidth]{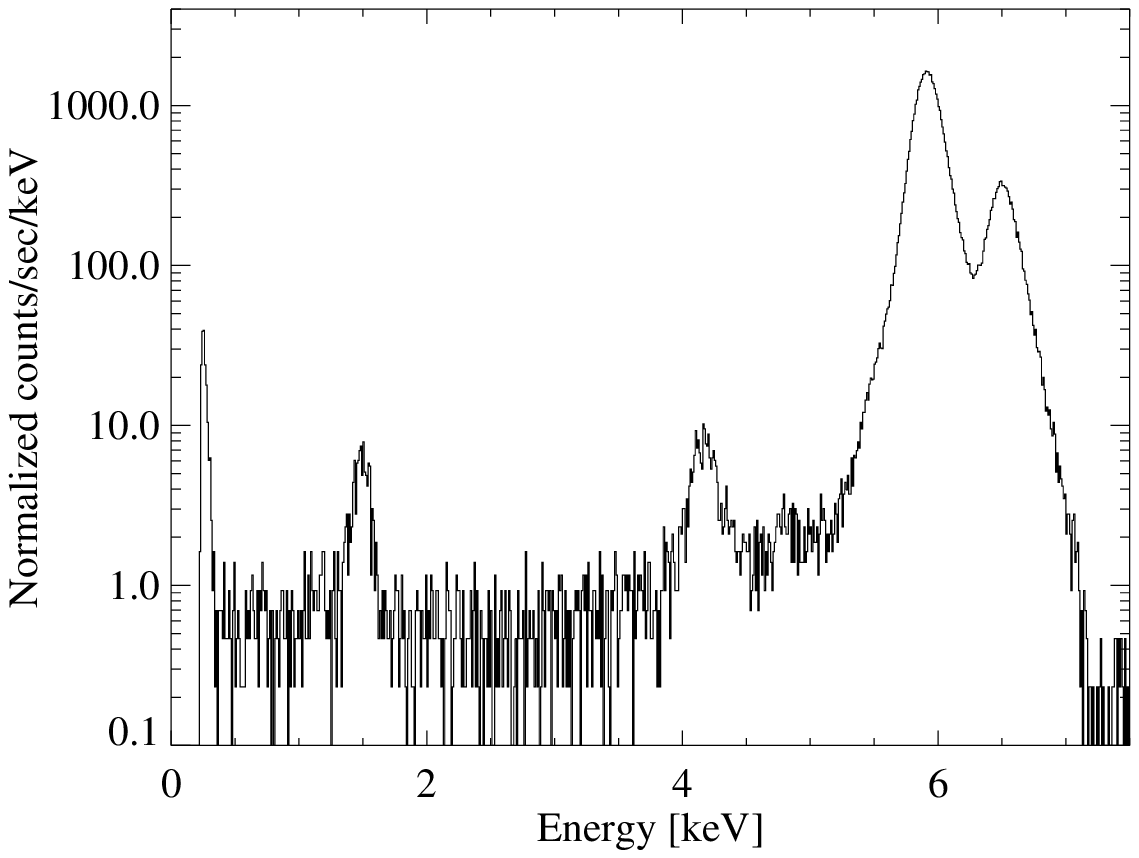}
  \end{minipage}
  \caption 
  {\label{fig:spectral-response} Left: Spectral energy response of an
    XMM-Newton EPIC pn-CCD similar to the one in use for CAST, for an
    incident monochromatic photon energy of 1.4\,\text{keV}
    \cite{popp:00a}. Right: Spectrum of an \fefifty calibration source
    measured with the CAST detector. The Al-K line apparent at
    $1.5\,\text{keV}$ originates from an aluminum filter installed in front
    of the \fefifty calibration source.}
\end{figure} 

\subsection{Spectral Distribution}
A time averaged background spectrum observed in 2004 under the same
operating conditions as during sun observations is shown in
Fig.~\ref{fig:bgrdspec}. The axion signal is expected as an excess signal
in the energy band between $1$ and $7\,\text{keV}$ where the background
spectrum has its minimum. The background level in this energy range
corresponds to a mean differential flux of
$(7.69\pm0.07)\times10^{-5}\,\text{counts}\,\text{cm}^{-2}\,\text{sec}^{-1}\,\text{keV}^{-1}$
which is equivalent to an integral background count rate of
$0.1\,\text{counts}\,\text{h}^{-1}$ in the focal spot with an area of
$6.4\,\text{mm}^{2}$. The features dominating the background spectrum are
fluorescent emission lines from Au, Cu, and Pb originating in the materials
close to the CCD chip on top of a Compton like continuum spectrum with a
slightly negative slope.  The Pb lines originate presumably form the
``ordinary solder'' that was used to fabricate the electronics board. This
solder can be contaminated by e.g. \pbtwoten. At energies below
$1\,\text{keV}$ a broad noise peak dominates the spectrum which is observed
in the CERN environment only. This background contribution seems to be
caused by the unusual high electronic noise level in the CAST area compared
to controlled laboratory conditions.  Since the building where the CAST
experimental area is located, also serves as an electronic support point
for the Large Hadron Collider--LHC and in general was not foreseen and
designed as an experimental area from the beginning, the electronic noise
level is unusual high and variable with time. This low energy background
component is limited to energies $E<0.8\,\text{keV}$, the region outside
the axion sensitive energy range, and it does not effect the sensitivity of
CAST.

\begin{figure}
  \centerline{\includegraphics[width=0.85\textwidth]{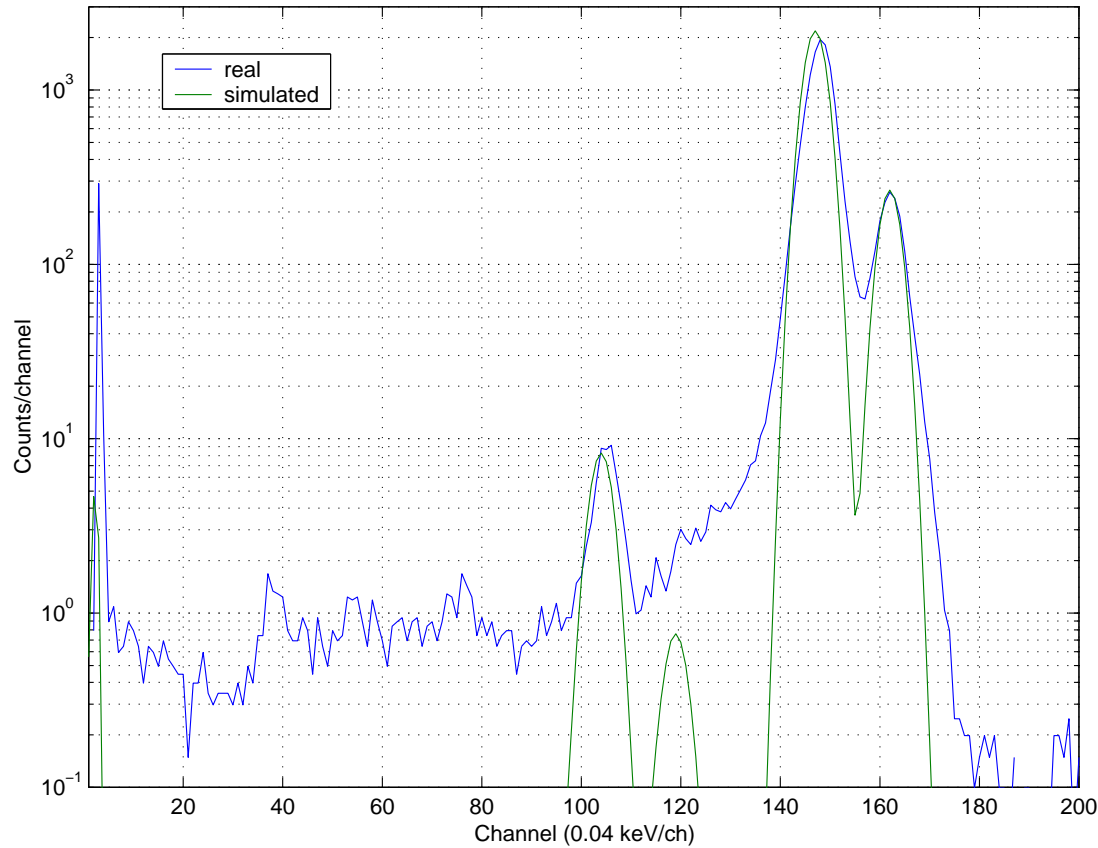}}
  \caption{\label{fig:sim-calspec} A simulated \fefifty calibration
    spectrum (green line) in comparison to a measured spectrum (blue line).
    The underground apparent between channel 10 and 140 is due to
    non-linearities in the detector response function (see
    Fig.~\ref{fig:spectral-response}).}
\end{figure} 

\subsection{Influence of the Detector Response Function}
Due to secondary energy loss effects in the semiconductor (partial-event
effect \cite{popp:00a}), the response of the pn-CCD detector to
mono-energetic photons is not a mere Gaussian, as expected from an ideal
detector with a finite energy resolution, but is slightly distorted for
incident photon energies with $E<6\,\text{keV}$. As an example, a spectrum
of monoenergetic X-rays with an energy of $1.4\,\text{keV}$ measured with
the pn-CCD detector is shown in Fig.~\ref{fig:spectral-response}. It is
apparent, that the Gaussian shaped main photo-peak is asymmetrically
distorted towards lower energies (``Shoulder'') and additional photons are
detected equally distributed between the noise peak and the shoulder of the
Gaussian main peak (``flat shelf''). The shape, height, and the width of
the shoulder as well as the flux level of the flat shelf have been
determined during the ground calibration campaign of the pn-CCD detector of
XMM-Newton. We refer the interested reader to
Ref.~\cite{popp:00a,haberl:02a} , for a more detailed introduction to the
partial-event effect and its implications on the spectral response of a
pn-CCD. The parameters that define the shape of the shoulder and the flat
shelf generally are energy dependant. Both effects, the asymmetric shoulder
and the flat shelf, are most distinct for incident photon energies $E <
6\,\text{keV}$ \cite{haberl:02a}. Due to this redistribution effect
especially photons in the flat shelf could contribute to the overall
background, if emission lines are present in the background spectrum. For
the CAST pn-CCD the ratio between hight of the Gaussian peak and the level
of the flat shelf (peak to valley ratio) is approximately 2700 for the
$\text{Mn-K}\alpha$ line as shown in the right part of
Fig.~\ref{fig:spectral-response}. Taking this ratio into account the
contribution of redistribution effects to the low energy background would
be two orders of magnitude below the actual level of sensitivity and
therefore is negligible.

\section{Background Simulations}
\begin{figure}
  \begin{center}
    \includegraphics[width=0.72\textwidth,angle=-90]{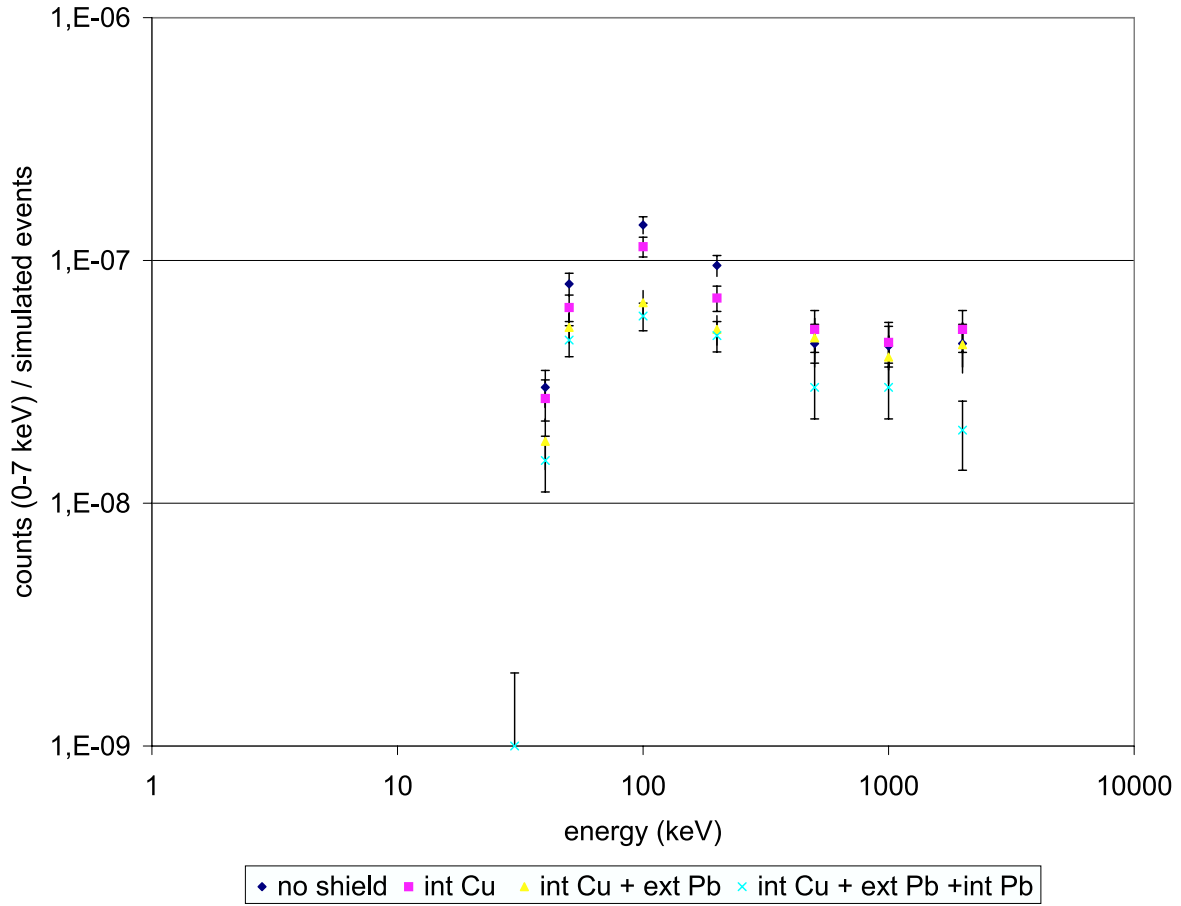}
  \end{center}
  \caption 
  {\label{fig:gamma-response} Response of the pn-CCD to external
    $\gamma$-ray background. The energy of the incident $\gamma$-ray photon
    versus the number of detected counts in the $0$--$7\,\text{keV}$ is
    shown relative to the number of simulated events for different
    configurations of the shield. Results are shown for an unshielded
    detector, the copper box (int Cu), an additional Pb shield outside the
    vessel (ext Pb), and an additional Pb shield inside the vessel being
    installed (int Pb).}
\end{figure} 
To understand the origin of the background observed with the CAST pn-CCD
and to quantify the relative contributions of different sources of
background we made extensive Monte-Carlo simulations with the GEANT4
package. As a first step, a calibration spectrum and the quantum efficiency
of the CCD were simulated to verify the reliability of the low energy
physical models implemented in GEANT4 and our simulation code. In general,
the response on a \fefifty source is very well reproduced including the
area ratio of the Si escape peak to the main emission peak (see
Fig.~\ref{fig:sim-calspec}). However, the simulated quantum efficiency in
the $10$--$15\,\text{keV}$ energy range is slightly higher, than the
results from calibration measurements.

\subsection{External $\gamma$-ray Background}
The energy dependant response of the pn-CCD to $\gamma$'s with an energy
$10\,\text{keV}<E<2\,\text{MeV}$ has been studied. For the simulations we
assumed a simplified spherical geometry with isotropic emission for the
incident external $\gamma$-ray background, and subsequently different
configurations of the pn-CCD passive shield have been taken into
consideration. The integral number of detected counts in the energy range
$0$--$7\,\text{keV}$ originating from high energy $\gamma$'s is shown in
Fig.~\ref{fig:gamma-response}. From our results it is obvious that
$\gamma$-rays with an energy close to $100\,\text{keV}$ contribute most to
the pn-CCD background and that the energy dependence of the contribution of
the external $\gamma$-ray background is low. The Pb shield suppresses
mainly $\gamma$'s with an energy above $100\,\text{keV}$, while the copper
shield has only a marginal effect on the $\gamma$ background.  From the
simulation we can derive a relative background reduction of a factor of
$1.6$--$2.5$ between an unshielded detector and the final shield
configuration of the CAST pn-CCD detector, which includes the internal Cu
box, the internal Pb shield, and the external Pb shield.  The simulations
underestimate the background reduction we have derived from background
measurements under different shielding configurations, which is a factor of
$3$ between an unshielded detector and the detector with its full shield
being installed. The reason of this discrepancy might be the
simplifications in the detector geometry we made, the uncertainties in the
response of the pn-CCD to high energy $\gamma$'s, and/or the isotropic
$\gamma$ emission geometry assumed for the simulations. In addition, the
background of internal sources in the detector set-up (natural
radioactivity) has been neglected in these simulations.

\subsection{External Neutron Background}
\begin{figure}
  \begin{center}
    \includegraphics[width=0.90\textwidth,angle=0]{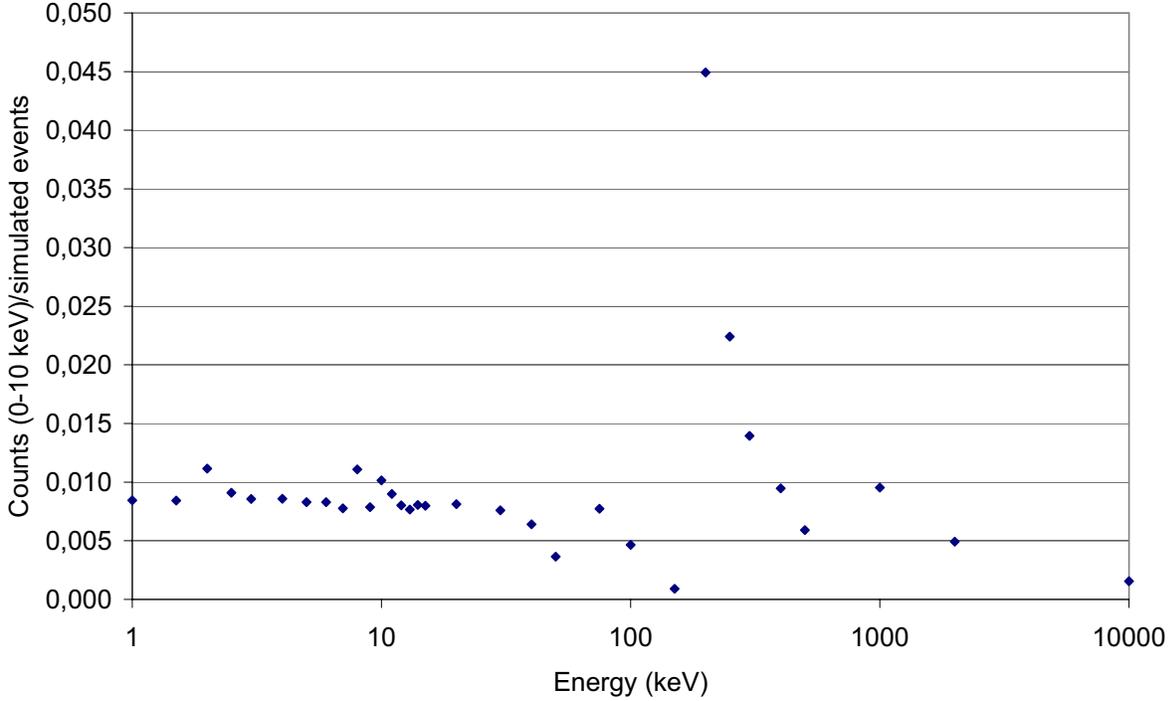}
  \end{center}
  \caption{\label{fig:neutron-response} Response of the pn-CCD to the neutron
    background.}
\end{figure}

Similar to the study for the $\gamma$-ray background, the influence of
thermal and high energy neutrons on the total background of the pn-CCD has
been evaluated. The energy of the incident neutrons has been varied from
$10^{-2}\,\text{eV}$ up to $10\,\text{MeV}$.  In
Fig.~\ref{fig:neutron-response} the ratio between events detected in the
$0$--$10\,\text{keV}$ energy range of visible energy (taking the quenching
factor into account) and the total number of simulated events as a function
of the incident energy of the neutron is shown. The process that dominates
the interaction of neutrons with the detector material seem to be elastic
scattering of neutrons off the silicon nuclei. The resonance visible in
Fig.~\ref{fig:neutron-response} corresponds to a resonance in the neutron
elastic scattering cross-section for silicon. Taking into account the
typical neutron flux at sea level\cite{heusser:95a,ziegler:98a} the
differential count rate expected in the pn-CCD detector from high energy
neutron interactions can be estimated to
$\approx6\times10^{-6}\,\text{counts}\,\text{cm}^{-2}\,\text{sec}^{-1}\,\text{keV}^{-1}$,
which is below the actual sensitivity of the detector. Although the
spectral distribution of the cosmic-ray induced neutron flux has a maximum
at thermal energies the estimated differential count rate of $3
\times{10}^{-8}\,\text{counts}\,\text{cm}^{-2}\,\text{s}^{-1}\,\text{keV}^{-1}$
originating in thermal neutron interactions is two orders of magnitude
smaller compared to the rates expected from high energy neutrons. Finally
we roughly estimated the muon-induced neutron background, which is around
two orders of magnitude lower than the flux of environmental neutrons.

\section{Conclusions and Prospects}
We have presented preliminary results from our study of the background of
the pn-CCD detector of CAST. Taking into account the results of our
simulations and background measurements we can conclude that effects due to
the non-linear detector response of the pn-CCD and neutron induced
background are negligible and do not contribute to the background of the
CAST pn-CCD detector at the present level of sensitivity. The major
contributions to the background seem to be environmental $\gamma$
background, internal natural radioactivity of the detector materials, and
muon induced background. In order to quantify these contributions, further
simulations based on a refined detector geometry are in progress.  Absolute
estimates of the background and the spectral shape of the background are
difficult to obtain, especially when the incident spectrum of, e.g.
environmental $\gamma$'s is not known. Measurements on environmental
background has been performed recently in the CAST experimental environment
and will further help to improve the simulations and modeling of the
background observed with the pn-CCD.

\nocite{lutz:04a}

%%%%%%%%%%%%%%%%%%%%%%%%%%%%%%%%%%%%%%%%%%%%%%%%%%%%%%%%%%%%%
\acknowledgments     %>>>> equivalent to \section*{ACKNOWLEDGMENTS}       
 
We acknowledge support by the Bundesministerium f\"ur Bildung und Forschung
(BMBF) under the grant number 05 CC2EEA/9, the N3 Dark Matter network of
the Integrated Large Infrastructure for Astroparticle Science -- ILIAS, and
the Virtuelles Institut f\"ur Dunkle Materie und Neutrinos -- VIDMAN. This
work has been performed within the CAST collaboration, we thank our
colleagues for their support.
%%%%%%%%%%%%%%%%%%%%%%%%%%%%%%%%%%%%%%%%%%%%%%%%%%%%%%%%%%%%%
%%%%% References %%%%%

\bibliography{mnemonic,xmm,cast,detback,conferences}   %>>>> bibliography data in report.bib
\bibliographystyle{spiebib}   %>>>> makes bibtex use spiebib.bst

\end{document}